\documentclass{aastex}
\usepackage{spr-astr-addons}

\begin{document}

\title{Reanalysis of two eclipsing binaries: \textbf{\emph{EE Aqr }}and \textbf{\emph{Z Vul}}}

\shorttitle{Short article title}
\shortauthors{Autors et al.}

\author{Sayyed Mohammad Reza Ghoreyshi\altaffilmark{1}}\affil{mo\_gh72@stu-mail.um.ac.ir}
\and \author{Jamshid Ghanbari\altaffilmark{1,2}}
\affil{ghanbari@ferdowsi.um.ac.ir} \and
\author{Fatemeh Salehi\altaffilmark{3}}
\affil{fsalehi@wali.um.ac.ir}

\altaffiltext{1}{ Department of Physics, School of Sciences,
Ferdowsi University of Mashhad, Mashhad,Iran\\}
\altaffiltext{2}{Department of Physics and Astronomy, San Francisco
State University, 1600 Holloway, San Francisco, CA 94132}
\altaffiltext{3}{Department of Physics, Khayyam Institute of Higher
Education, Mashhad, Iran}


\begin{abstract}
We study the radial-velocity and light curves of the two eclipsing
binaries $EE$ $Aqr$ and $Z$ $Vul$. Using the latest version of the
Wilson \& Van Hamme (2003) model, absolute parameters for the
systems are determined. We find that $EE$ $Aqr$ and $Z$ $Vul$ are
near-contact and semi-detached systems, respectively. The primary
component of $EE$ $Aqr$ fills about 96\% of its 'Roche lobe', while
its secondary one appears close to completely filling this limiting
volume. In a similar way, we find fill-out proportions of about 72
and 100\% of these volumes for the primary and secondary components
of $Z$ $Vul$ respectively. We compare our results with those of
previous authors.
\end{abstract}



\section{Introduction}

Eclipsing binary stars studies often involve the combination of
photometric (light curve) and spectroscopic (mainly, radial velocity
curve) data. The analysis of the light-velocity curves enable
astronomers to obtain absolute physical parameters describing the
system and its components. The physical parameters derived from the
photometric and spectroscopic data can help to improve our
understanding of physical processes in stars. In view of the
importance of $EB$-$EA$ type binary stars in the problem of mass
exchange in binaries and in the theory of their evolution, the study
of these systems play an important role in finding of the complex
initial phases of stellar evolution. In order to understand the
physical nature of $EB$ and $EA$ type binaries, two candidates, $EE$
$Aqr$ and $Z$ $Vul$, were selected based on the following criteria:
a) the shape of the light curve, b) their photoelectric measurements
are not enough and also somewhat the analysis results derived by the
previous authors appear to differ from each other. The first, we
state their history in literature and then describe our solution
method.

\subsection{$EE$ $Aqr$}

The binary system $EE$ $Aqr$ (HD213863, BD-$20^\circ$6454, SAO
191236, $\alpha_{2000} = 22^h34^m41^s.87,$ $\delta_{2000} =
19^\circ51'34''$) was discovered to be variable by Strohmeier \&
Knigge (1960). Strohmeier et al. (1962) derived its eclipsing
behavior with a orbital period of $0.51$ $d$ and classified it as an
Algol (EA) type eclipsing binary. Later, the photoelectric
observations were obtained by Williamon (1974), Padalia (1979) and
Covino et al. (1988). Williamon (1974) who showed the light curves
to be more like that of $\beta$ Lyrae (EB) type with a period of
$0.5089951$ $d$ and analyzed these using the method of Russell \&
Merrill (1952). Thus preliminary photometric orbital parameters of
the system were derived. Slightly different orbital parameters were
determined by Padalia (1979) for the system using the same method.
Russo \& Sollazzo (1982) found many inconsistencies in applying the
Russell and Merrill (1952) method of solution. Therefore they
reanalyzed both $V$ light curves of Williamon (1974) and Padalia
(1979) using the Wilson \& Devinney (1971) model. They obtained the
same results for both light curves while the Russell-Merrill model
does not. They found a solution as a semi-detached system. The
system has also been studied spectroscopically by Hilditch \& King
(1988), from which radial-velocities were measured for both
components by means of the cross-correlation code VCROSS. The first
simultaneous solution of photometric and radial-velocity curves was
performed by Covino et al. (1990) based on the Wilson \& Devinney
(1971) model. Their solution indicated that $EE$ $Aqr$ might be a
contact configuration not yet in thermal equilibrium. From 1990 to
now, no analysis of light curves and new photometric observations
has been reported.

The period variations of the system has studied by Srivastava (1987)
who found irregular changes of the order of $\pm10^{-6}$ $d$ and
many observations of the times of minima (Mallama (1980), Covino et
al. (1988), Deeg et al.(2003)) so far.

The spectral classes of the components of $EE$ $Aqr$ were given as
F0 (Williamon (1974)), FV+A (Padalia (1979)), A8V+(K3-K4)(Russo \&
Sollazzo (1982)).

\subsection{$Z$ $Vul$}

The variability of $Z$ $Vul$ (HD181987, $\alpha_{2000} =
19^h\;21^m\\*39^s.11,$ $\delta_{2000} = 25^\circ34'29''.45$) was
discovered to be an eclipsing binary by Herschel (It has been
reported by Astbury, 1909). Plaskett (1920), Petrie (1950), Roman
(1956), Popper (1957b) and Cester et al. (1977) determined the
spectral type of the components as B3, B4+B6, B5V+A, B3-4V+A2-3III
and B2V+A1III respectively. The first researches on the system
presented its eclipsing behavior with a orbital period slightly less
than $2.5$ $d$ and classified it as an Algol (EA) type eclipsing
binary. Cester et al. (1977) analyzed Broglia's photometric
observations (1964) using the $WINK$ program (Wood, 1972) and
determined slightly different masses, radii and luminosities for the
system with those found in the literature. They found a solution as
a semi-detached system.

Peters (1994) presented the first IUE observations to investigate
the effect of the cross-section and temperature of the primary on
the mass transfer and mass loss in the system. They found that the
primary involve with weak winds, originating from the gas stream
will strike the primary instead of forming an accretion disk. The
lack of H-alpha emission in system supports this proposition.

The first results from far-UV observations obtained by Peters \&
Polidan (1997) about the nature of the circum stellar material in
the system. Their observations showed no infalling from a gas
stream. So it is consistent with this fact that the system is
sufficiently close and can not establish a disk.

It is recognized that depend on applying present models, we can
obtained the same or different results of analyzing light-velocity
curves and thus it follows an low or high understanding of stellar
evolution. Therefor it is need to research a tool for a better
comprehension of the binaries structure and evolutive stage.

In this study the authors adopted the more realistic close binary
model based on the Roche model. In order to obtain accurate
solutions in this situation, the light curves were analyzed by means
of the latest the Wilson's computer code. The computing code used
was developed by the Wilson \& Van Hamme code (2003)(here after
\textbf{WV}) to determine the parameters of the system. We selected
it as our analysis research tool both for its intrinsic virtues and
because of some improvements in comparison with earlier code (Wilson
\& Divinney, 1971). The fourth revision (2003) is improved for
example based on Kurucz's new atmospheres, log g as a parameter
(allowing for handling giants, sub-giants, etc., in addition to main
sequence stars) so that temperature ranges vary according to log g
together with 19 abundances (relative to the sun). Hence we felt it
useful to reanalyze these systems and get the most reliable elements
with the model of \textbf{WV}, which was developed especially to
handel close systems and compare our finding with results of the
previous authors.

All the data we analyze in the paper are taken from the literature
so that these data appeared to be free from any peculiarities
reported by the previous researches. We consider photometric
observations of Williamon (1974) (here after \textbf{WI}) and
spectroscopic observations of Hilditch \& King (1988) (here after
\textbf{HK}) for $EE$ $Aqr$ and the photometric observations of
Broglia (1964) (here after \textbf{BR}) and spectroscopic
observations of Popper (1957b) (here after \textbf{PO}) for $Z$
$Vul$.

The present paper is organized as follows. The assumptions are
described in section 2. Photometric solutions of light curves is
given in section 3. Spectroscopic solutions of radial velocities
curves in section 4. Absolute elements for the primary and secondary
components in section 5 and in section 6, we give conclusion.


\section{Assumptions}

The latest 2003 version of the Wilson program was applied for
photometric and spectroscopic solutions. The method assumes the star
surfaces to be equipotential and uses a set of curve-dependent or
curve-independent parameters that can be adjusted by $LC$ and $DC$
programs: the orbital inclination $i$, surface potentials
$\Omega_{1,2}$, the mean surface effective temperatures $T_{1,2}$,
the mass ratio $q=m_2/m_1$, the bandpass luminosities $L_{1,2}$, the
wavelength-specific limb darkening coefficients $x_{1,2}$, the
bolometric limb darkening coefficients $x_{1,2}(bol)$, the
bolometric gravity darkening exponents g$_{1,2}$ and the bolometric
albedos $A_{1,2}$. Throughout this paper, the subscripts 1 and 2
refer to the primary (hotter) and the secondary (cooler) components,
respectively.

For both components of the systems, we used bolometric linear,
logarithmic and square root law and the best result was obtained for
bolometric logaritmic limb darkening law of Klinglesmith \& Sobieski
(1970) of the form:

\begin{equation}
I=I_\circ(1-x+x \cos\theta-y \cos\theta \ln(\cos\theta)),
\end{equation}
where the limb darkening coefficients $x$ and $y$ for both
components were fixed to their theoretical values, interpolated
using Van Hamme's (1993) formula which have tabulated in tables 1
and 2.

The gravity darkening exponent from Lucy (1967) and the bolometric
albedos from Rucinski (2001) were chosen for convective envelopes
($g=0.32, A=0.5$) and for radiative envelopes ($g=1.0, A=1.0$),
which are agreement with the final surface temperature. In order to
reduce the number of free parameters these parameters were kept
constant during all the iterations. Also it is assumed that this
binary system has zero orbital eccentricity ($e = 0.0$) and that its
rotational and orbital spins are synchronous ($F_1 = F_2 = 1.0$).
Also black body models are employed, and we assume that there is no
third light ($l_3=0.0$) for both systems.

\begin{table}[t]
\caption{The limb darkening coefficients for $EE$ $Aqr$}
\begin{tabular}{@{}llll@{}}
\hline
Parameters & Filter B & Filter U & Filter V \\
\hline
$x_1(bol)$ & 0.640 & 0.640 & 0.640 \\
$y_1(bol)$ & 0.255 & 0.255 & 0.255 \\
$y_2(bol)$ & 0.153 & 0.153 & 0.153 \\
$x_1$ & 0.690 & 0.795 & 0.788 \\
$x_2$ & 0.800 & 0.802 & 0.834 \\
$y_1$ & 0.291 & 0.328 & 0.279 \\
$y_2$ & -0.014 & -0.411 & -0.189 \\
\hline
\end{tabular}
\end{table}

\begin{table}[t]
\caption{The limb darkening coefficients for $Z$ $Vul$}
\begin{tabular}{@{}llll@{}}
\hline
Parameters & Filter B & Filter U & Filter V \\
\hline
$x_1(bol)$ & 0.762 & 0.762 & 0.762 \\
$x_2(bol)$ & 0.703 & 0.703 & 0.703 \\
$y_1(bol)$ & 0.090 & 0.090 & 0.090 \\
$y_2(bol)$ & 0.072 & 0.072 & 0.072 \\
$x_1$ & 0.434 & 0.482 & 0.509 \\
$x_2$ & 0.584 & 0.615 & 0.682 \\
$y_1$ & 0.233 & 0.228 & 0.275 \\
$y_2$ & 0.290 & 0.249 & 0.336 \\
\hline
\end{tabular}
\end{table}


\section{Photometric Solutions}

Using the photometric data from the \textbf{WI} and \textbf{BR}, we
derived the final elements of the systems. Before beginning the
analysis, we have chosen some parameters of the systems using the
spectroscopic and photometric information as a starting input
values.

Initially, the light curve program (LC) was implemented in mode 2
with no third light (based on the spectroscopic observations,
\textbf{HK} and \textbf{PO}) corresponding to detached configuration
by choosing i, $T_2$ (while the temperature of the primary component
was assumed from spectroscopic or color data), $\Omega_{1,2}$ and
$L_1$ as adjustable parameters in UBV filters. The symmetry between
the maximums in the light curves indicate the lack of star spots on
the components. After a few runs of the LC program an initial set of
values was employed as input parameters for the DC program. After a
few runs with the differential correction program, we could not
obtain the values of the physical parameters with less error, thus
making change to mode 5 so that expected from the earlier workers.
By several iterations of the $LC$ and $DC$ programs and adjusting
the parameters $i,T_2,\Omega_1$ and $L_1$ for each filter, finally
the solution evolved into a nearcontact configuration for $EE$ $Aqr$
and semi-detached one for $Z$ $Vul$. The results of light curve
solutions with the final elements are given in Tables 3 and 4. The
theoretical light curves computed with these results are shown in
Figures 1 and 2. The agreement between the observed (solid circles)
and the theoretical light curves (continuous lines) is quite good.

\begin{table*}
\small \caption{The photometric parameters of $EE$ $Aqr$}
\begin{tabular}{@{}llllllllll@{}}
\hline
Parameters & This work & This work & This work & Williamon & Padalia & Russo and & Hilditch & Covino et & Covino et\\
& B Filter & U Filter & V Filter & (1974) & (1979) & Sollazzo & and King & al. (1990) & al. (1990) \\
& (Johnson) & (Johnson) & (Johnson) & & & (1982) & (1988) & $BVRI$ & $BVRI$\&$RV$ \\
\hline & & & & & & & & & \\
$i$ & 80.5074 & 81.3365 & 78.9509 & 73.71 & 74.00 & 76.0 & 80.5 & 80.0 & 80.2 \\
& $\pm$ 0.2975 & $\pm$ 0.4360 & $\pm$ 0.2379 & & & & & & \\
$q$ & 0.32 & 0.32 & 0.32 & 0.503 & 0.500 & 0.4 & 0.32 & 0.332 & 0.327 \\
& fixed & fixed & fixed & & & & & & \\
$T_1$ & 7060 & 7060 & 7060 & 8000 & 8000 & 8000 & 7060 & 7230 & 7227 \\
& fixed & fixed & fixed & & & & & & \\
$T_2$ & 4200 & 3952 & 4173 & 4340 & 4365 & 4440 & 4395 & 4240 & 4233 \\
& $\pm$ 116 & $\pm$ 252 & $\pm$ 85 & & & & & & \\
$\Omega_1$ & 2.6178 & 2.6141 & 2.6030 & 3.002 & 3.003 & 2.78 & & 2.619 & 2.568 \\
& $\pm$ 0.0118 & $\pm$ 0.0106 & $\pm$ 0.0116 & & & & & & \\
$\Omega_2$ & 2.5100 & 2.5100 & 2.5100 & 2.882 & 2.875 & 2.68 & & 2.538 & 2.561 \\
& & & & & & & & & \\
$L_1/(L_1+L_2)$ & 0.9832 & 0.9950 & 0.9716 & 0.966 & 0.964 & 0.951 & 0.952 & 0.959 & \\
& & & & & & & & & \\
$L_2/(L_1+L_2)$ & 0.0168 & 0.0050 & 0.0284 & 0.034 & 0.036 & 0.049 & 0.048 & 0.041& \\
& & & & & & & & & \\
log g$_1$(CGS)& 4.33 & 4.33 & 4.33 & & & & & & \\
& & & & & & & & & \\
log g$_2$(CGS) & 4.24 & 4.24 & 4.24 & & & & & & \\
& & & & & & & & & \\
$r_1(pole)$ & 0.4303 & 0.4310 & 0.4330 & 0.394 & 0.394 & & 0.434 & 0.43 & 0.44 \\
& $\pm$ 0.0021 & $\pm$ 0.0019 & $\pm$ 0.0021 & & & & & & \\
$r_1(point)$ & 0.5121 & 0.5139 & 0.5194 & 0.469 & 0.467 &  & & 0.52 & 0.54 \\
& $\pm$ 0.0056 & $\pm$ 0.0051 & $\pm$ 0.0059 & & & & & & \\
$r_1(side)$ & 0.4568 & 0.4577 & 0.4603 & 0.415 & 0.414 & & & 0.46 & 0.47 \\
& $\pm$ 0.0027 & $\pm$ 0.0025 & $\pm$ 0.0027 & & & & & & \\
$r_1(back)$ & 0.4765 & 0.4775 & 0.4806 & 0.437 & 0.435 & & & 0.48 & 0.49 \\
& $\pm$ 0.0033 & $\pm$ 0.0030 & $\pm$ 0.0033 & & & & & & \\
$r_2(pole)$ & 0.2659 & 0.2659 & 0.2659 & 0.300 & 0.299 & & 0.262 & 0.27 & 0.26 \\
& $\pm$ 0.0025 & $\pm$ 0.0017 & $\pm$ 0.0032 & & & & & & \\
$r_2(point)$ & 0.3853 & 0.3853 & 0.3853 & 0.406 & 0.405 & & & 0.36 & 0.33 \\
& $\pm$ 0.0121 & $\pm$ 0.0080 & $\pm$ 0.0153 & & & & & & \\
$r_2(side)$ & 0.2769 & 0.2769 & 0.2769 & 0.313 & 0.312 & & & 0.28 & 0.27 \\
& $\pm$ 0.0027 & $\pm$ 0.0018 & $\pm$ 0.0034 & & & & & & \\
$r_2(back)$ & 0.3096 & 0.3096 & 0.3096 & 0.345 & 0.345 & & & 0.31 & 0.30 \\
& $\pm$ 0.0027 & $\pm$ 0.0018 & $\pm$ 0.0034 & & & & & & \\
$\Omega_{in}$ & 2.5100 & 2.5100 & 2.5100 & & & & & & \\
& & & & & & & & & \\
$\Omega_{out}$ & 2.3114 & 2.3114 & 2.3114 & & & & & & \\
& & & & & & & & & \\
$\Sigma$ $\omega(o-c)^2$ & 0.0172 & 0.0234 & 0.0156 & & & & & & \\
& & & & & & & & & \\
\hline
\end{tabular}
\end{table*}

\begin{table*}
\caption{The photometric parameters of $Z$ $Vul$}
\begin{tabular}{@{}llllllll@{}}
\hline Parameters & This work & This work & This work & Popper &
\multicolumn{3}{@{}c@{}}{Cester et al. (1977)}
\\ \cline{6-1}\cline{7-1}\cline{8-1}
& B Filter & U Filter & V Filter & (1957b) & B Filter & U Filter & V Filter \\
& (Johnson) & (Johnson) & (Johnson) & & (Johnson) & (Johnson) & (Johnson) \\
\hline & & & & & & & \\
$i$ & 88.6106 & 88.5185 & 88.6478 & 88 & 88 & 88.5 & 88.9 \\
& $\pm$ 0.0666 & $\pm$ 0.0453 & $\pm$ 0.0772 & & & & \\
$q$ & 0.43 & 0.43 & 0.43 & & 0.43 & 0.43 & 0.43\\
& fixed & fixed & fixed & & & & \\
$T_1$ & 19840 & 19840 & 19840 & & 19850 & 19850 & 19840 \\
& fixed & fixed & fixed & & & & \\
$T_2$ & 10909 & 9882 & 10810 & & 9000 & 10290 & 9410 \\
& $\pm$ 25 & $\pm$ 32 & $\pm$ 27 & & & & \\
$\Omega_1$ & 3.7975 & 3.8512 & 3.7434 & & & & \\
& $\pm$ 0.0106 & $\pm$ 0.0083 & $\pm$ 0.0119 & & & & \\
$\Omega_2$ & 2.7387 & 2.7387 & 2.7387 & & & & \\
& & & & & & & \\
$L_1/(L_1+L_2)$ & 0.8151 & 0.8869 & 0.7889 & & 0.962 & 0.930 & 0.954 \\
& & & & & & & \\
$L_2/(L_1+L_2)$ & 0.1849 & 0.1131 & 0.2111 & & 0.038 & 0.070 & 0.046 \\
& & & & & & & \\
log g$_1$(CGS)& 3.87 & 3.89 & 3.86 & & & & \\
& & & & & & & \\
log g$_2$(CGS) & 3.49 & 3.49 & 3.49 & & & & \\
& & & & & & & \\
$r_1(pole)$ & 0.2954 & 0.2908 & 0.3002 & & & & \\
& $\pm$ 0.0009 & $\pm$ 0.0007 & $\pm$ 0.0011 & & & & \\
$r_1(point)$ & 0.3086 & 0.3031 & 0.3144 & & & & \\
& $\pm$ 0.0011 & $\pm$ 0.0008 & $\pm$ 0.0013 & & & & \\
$r_1(side)$ & 0.3011 & 0.2962 & 0.3063 & & & & \\
& $\pm$ 0.0010 & $\pm$ 0.0007 & $\pm$ 0.0012 & & & & \\
$r_1(back)$ & 0.3058 & 0.3006 & 0.3113 & & & & \\
& $\pm$ 0.0011 & $\pm$ 0.0008 & $\pm$ 0.0012 & & & & \\
$r_2(pole)$ & 0.2881 & 0.2881 & 0.2881 & & & & \\
& $\pm$ 0.0004 & $\pm$ 0.0004 & $\pm$ 0.0005 & & & & \\
$r_2(point)$ & 0.4142 & 0.4142 & 0.4142 & & & & \\
& $\pm$ 0.0018 & $\pm$ 0.0016 & $\pm$ 0.0021 & & & & \\
$r_2(side)$ & 0.3004 & 0.3004 & 0.3004 & & & & \\
& $\pm$ 0.0004 & $\pm$ 0.0004 & $\pm$ 0.0005 & & & & \\
$r_2(back)$ & 0.3330 & 0.3330 & 0.3330 & & & & \\
& $\pm$ 0.0004 & $\pm$ 0.0004 & $\pm$ 0.0005 & & & & \\
$\Omega_{in}$ & 2.7387 & 2.7387 & 2.7387 & & & & \\
& & & & & & & \\
$\Omega_{out}$ & 2.4781 & 2.4781 & 2.4781 & & & & \\
& & & & & & & \\
$\Sigma$ $\omega(o-c)^2$ & 0.0039 & 0.0057 & 0.0031 & & & & \\
& & & & & & & \\
\hline
\end{tabular}
\end{table*}

\begin{figure}[t]
\includegraphics[width=0.48\textwidth]{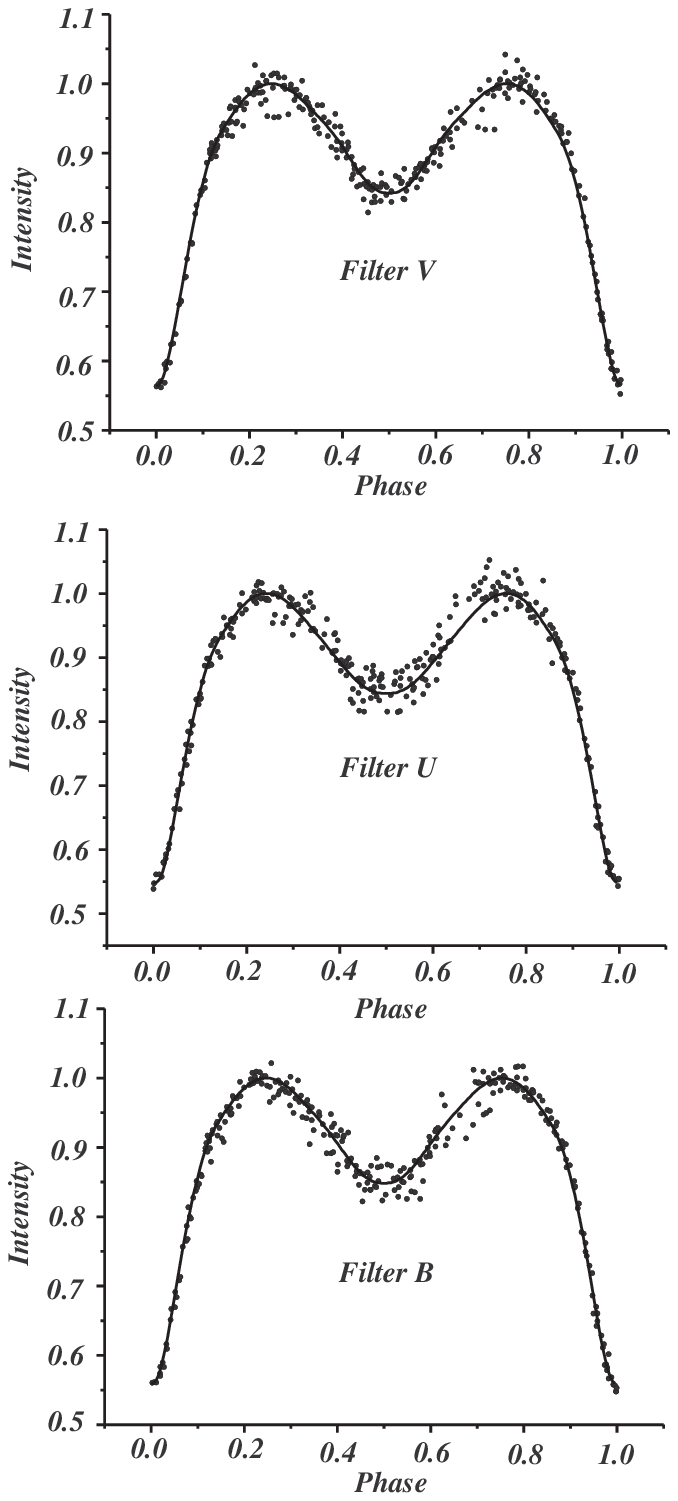}
\caption{The observed and theoretical light curves of $EE$ $Aqr$.
Solid circles show the observed data and the theoretical light
curves are shown by continuous lines.}
\end{figure}

\begin{figure}[t]
\includegraphics[width=0.48\textwidth]{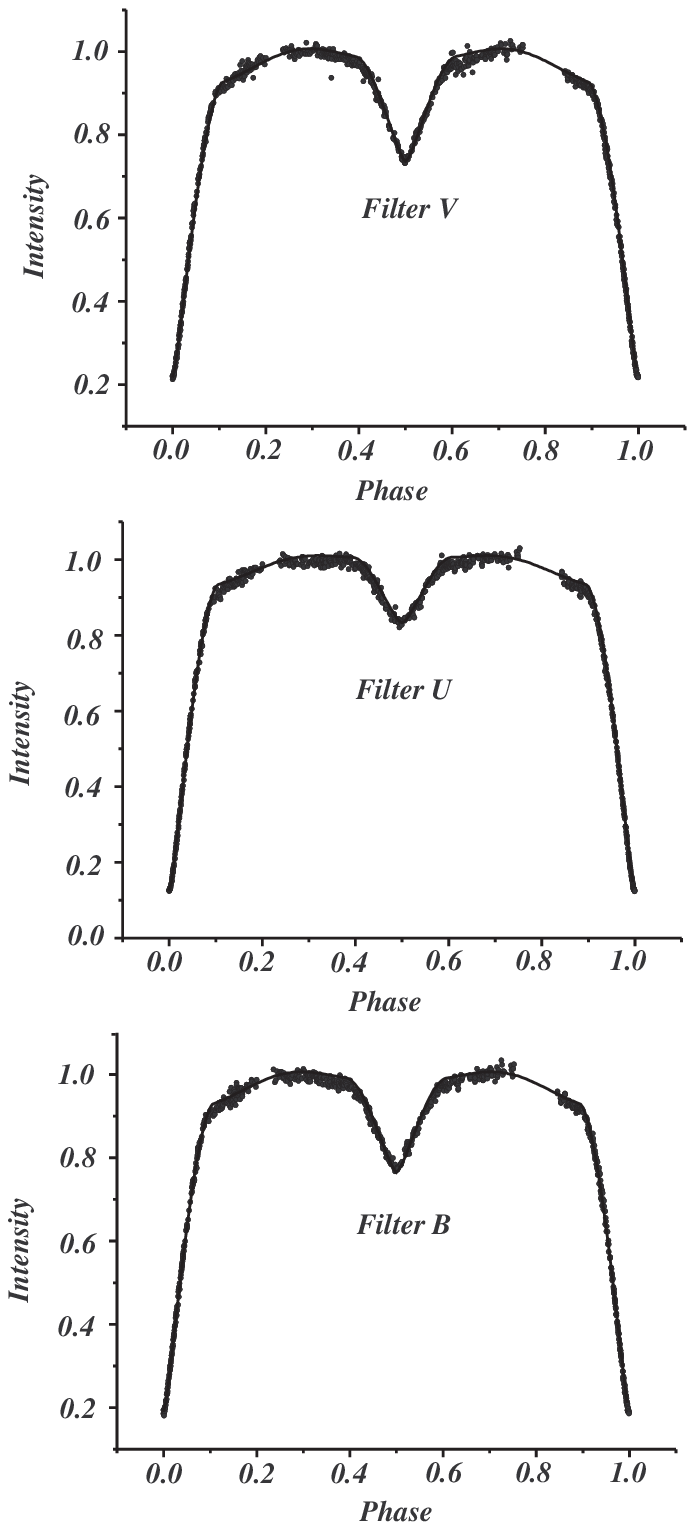}
\caption{The observed and theoretical light curves of $Z$ $Vul$.
Solid circles show the observed data and the theoretical light
curves are shown by continuous lines.}
\end{figure}

\section{Spectroscopic Solutions}

Using the radial-velocity data from \textbf{HK} and \textbf{PO}, we
derived the orbital elements of the systems. Adjustable parameters
were the following: the orbital semi-major axis $a$, the radial
velocity of the binary system center of mass $V_\gamma$. The results
are given in table 5. Using the final elements of the objects, the
theoretical radial-velocity curves are shown in Figure 3. The
agreement between the observed (solid circles) and the theoretical
radial-velocity curves (continuous lines) is quite good.

\begin{table}[t]
\caption{The spectroscopic parameters}
\begin{tabular}{@{}lll@{}}
\hline
Parameters & $EE$ $Aqr$ & $Z$ $Vul$\\
\hline
$V_\gamma (Km/s)$ & -1.6786 & -22.7690  \\
& $\pm$ 0.6520 & $\pm$ 0.3952  \\
\hline
$a (R_\odot)$ & 4.1240 & 15.8411 \\
& $\pm$ 0.0188 & $\pm$ 0.0454 \\
\hline
\end{tabular}
\end{table}


\begin{figure}[t]
\includegraphics[width=0.48\textwidth]{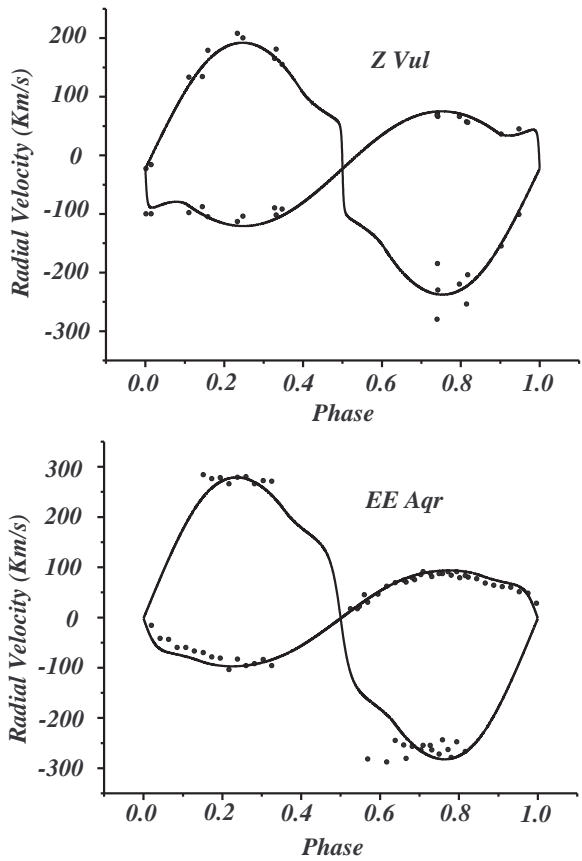}
\caption{The observed and theoretical radial-velocity curves of $EE$
$Aqr$ and $Z$ $Vul$. Solid circles show the observed data and the
theoretical radial-velocity curves are shown by continuous lines.}
\end{figure}


\begin{figure}[t]
\includegraphics[width=0.48\textwidth]{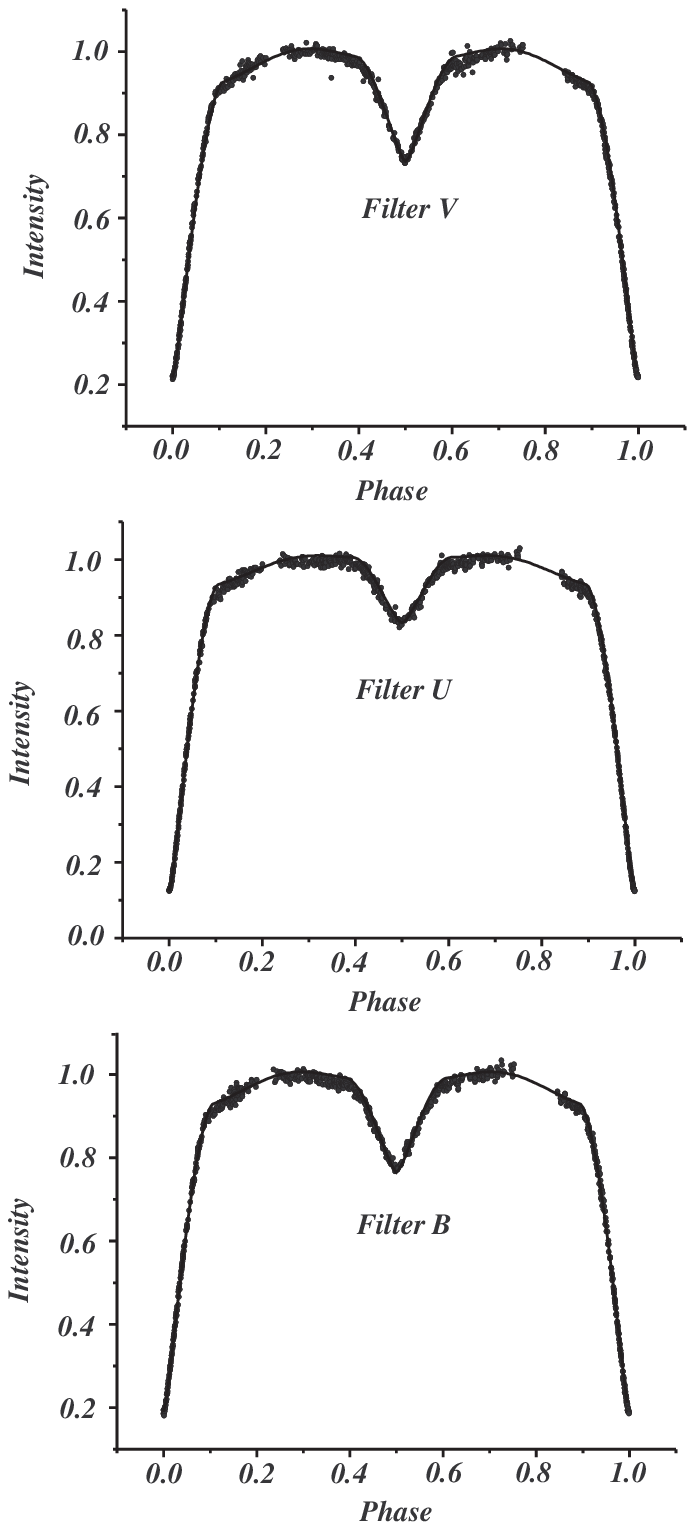}
\caption{The observed and theoretical radial-velocity curves of
\emph{AV Del}. Solid circles show the observed data and the
theoretical radial-velocity curve is shown by continuous lines.}
\end{figure}


\section{Absolute Elements}

Using the obtained results of the light and radial-velocity curves,
we calculate the absolute parameters of the systems. The results are
listed in tables 6 and 7. We determine the absolute dimensions using
the following formulae:

\begin{equation}
M_1/M_\odot=f_1(M_1,M_2,i)(1+q)^2/\sin^3i,
\end{equation}
\begin{equation}
f_1(M_1,M_2,i)=(M_2\sin i)^3/(M_1+M_2)^2,
\end{equation}
\begin{equation}
M_2/M_\odot=q(M_1/M_\odot),
\end{equation}
\begin{equation}
R_{1,2}/R_\odot=4.207(M_{1,2}/M_\odot(1+q)P^2(days))^{1/3}r(side),
\end{equation}
\begin{equation}
L_{1,2}/L_\odot=(R_{1,2}/R_\odot)^2(T_{1,2}/T_\odot)^4,
\end{equation}
\begin{equation}
\rho_{1,2}/\rho_\odot=(0.01344M_{1,2})/[(M_1+M_2)P^2r^3_{1,2}],
\end{equation}
\begin{equation}
(M_{bol})_{1,2}=42.36-10\log T_{1,2}-5\log (R_{1,2}/R_\odot).
\end{equation}

\begin{table*}
\caption{The absolute elements of the binary system $EE$ $Aqr$}
\begin{tabular}{@{}lllllll@{}}
\hline
Parameters & This Work & This Work & This Work & Russo and & Hilditch and & Covino et \\
& Filter B & Filter U & Filter V & Sollazzo & King (1988) & al. (1990) \\
& (Johnson) & (Johnson) & (Johnson) & (1982) & & \\
\hline
$M_1/M_\odot$ & 2.761 & 2.761 & 2.761 & 1.9 & 2.2 & 2.14 \\
$M_2/M_\odot$ & 0.883 & 0.883 & 0.883 & 0.95 & 0.71 &  0.70 \\
$L_1/L_\odot$ & 6.55 & 6.55 & 6.55 & 9.16 & & 7.8 \\
$L_2/L_\odot$ & 0.32 & 0.25 & 0.32 & 0.47 & & 0.33 \\
$R_1/R_\odot$ & 1.88 & 1.88 & 1.88 & 1.58 & 1.75 & 1.79 \\
$R_2/R_\odot$ & 1.18 & 1.18 & 1.18 & 1.21 & 1.07 & 1.06 \\
$\rho_1/\rho_\odot$ & 0.4123 & 0.4099 & 0.4030 & & & \\
$\rho_2/\rho_\odot$ & 0.5922 & 0.5922 & 0.5922 & & & \\
$M_1(bol)$ & 2.55 & 2.55 & 2.53 & & & \\
$M_2(bol)$ & 5.82 & 6.08 & 5.85 & & & \\
$f_1(M_1,M_2,i)$ & 0.0498 & 0.0501 & 0.0490 & & & \\
\hline
\end{tabular}
\end{table*}

\begin{table*}
\caption{The absolute elements of the binary system $Z$ $Vul$}
\begin{tabular}{@{}llllll@{}}
\hline
Parameters & This Work & This Work & This Work & Popper (1957b) & Cester et al. (1977) \\
& Filter B & Filter U & Filter V & & \\
& (Johnson) & (Johnson) & (Johnson) & & \\
\hline
$M_1/M_\odot$ & 6.209 & 6.209 & 6.209 & 5.4 & 5.4 \\
$M_2/M_\odot$ & 2.670 & 2.670 & 2.670 & 2.3 & 2.3 \\
$L_1/L_\odot$ & 2631.37 & 2543.85 & 2720.37 & 1850 & 2818 \\
$L_2/L_\odot$ & 252.75 & 170.21 & 243.75 & 175 & 162 \\
$R_1/R_\odot$ & 4.77 & 4.69 & 4.85 & 4.7 & 4.5 \\
$R_2/R_\odot$ & 4.89 & 4.89 & 4.89 & 4.7 & 4.6 \\
$\rho_1/\rho_\odot$ & 0.0571 & 0.0600 & 0.0543 & & \\
$\rho_2/\rho_\odot$ & 0.0247 & 0.0247 & 0.0247 & & \\
$M_1(bol)$ & -3.96 & -3.93 & -4.00 & -3.5 & \\
$M_2(bol)$ & -1.42 & -0.99 & -1.38 & -1.0 & \\
$f_1(M_1,M_2,i)$ & 0.2412 & 0.2412 & 0.2412 & & \\
\hline
\end{tabular}
\end{table*}

\section{Conclusion}

Comparing the new photometric-spectroscopic solutions of the systems
with those of given in literature, the present authors can get
following conclusions:

1-In our study, we may conclude that the systems are near-contact
and semi-detached  for $EE$ $Aqr$ and $Z$ $Vul$ respectively. The
fillout for the components can be calculated from the following
formula:

\begin{equation}
fillout_{1,2}=\frac{\Omega_{in}}{\Omega_{1,2}}\times100.
\end{equation}

\smallskip
Also from the following formulae we calculate $K_{1,2}$ and
$a_{1,2}\sin i$ for each of the systems:

\begin{equation}
K_{1,2}=\frac{2\pi a_{1,2}\sin i}{P},
\end{equation}
\begin{equation}
a_1\sin i=[\frac{GP^2f_1(M_1,M_2,i)}{4\pi^2}]^\frac{1}{3},
\end{equation}
\begin{equation}
a_2\sin i=\frac{a_1\sin i}{q}.
\end{equation}

The derived values of the parameters are given in tables 8 and 9.
According to our results, $EE$ $Aqr$ is a near-contact system which
the primary and secondary components filling are almost $96\pm1$ and
$100$ percent of their respective critical Roche lobes. Also $Z$
$Vul$ is a semi-detached system that the primary and secondary
components filling are almost $72\pm1$ and $100$ percent of their
respective critical Roche lobes. The fillout percentage was not
accurately determined in previous workers ($86\pm3$ and $94\pm3$
percent according to \textbf{HK} for the primary and secondary
components of $EE$ $Aqr$, respectively; no information for ones of
$Z$ $Vul$). It seems that fillout obtained from our solutions is
appropriate for highly evolved system.

2- According to our synthetic light curves, $EE$ $Aqr$ is
\emph{$\beta$-Lyrae} type and $Z$ $Vul$ is an $Algol$ type system.
Also according to the obtained parameters of the systems, we have
drawn the configuration of the components using the Binary Maker 2.0
(BradStreet, 1993) software, which are shown in figures 4 and 5.

3- The new values of the velocity amplitudes and the gama velocity
differ only slightly from earlier researches.

4- The absolute elements of both components determine the
evolutionary state of the systems. Entering the results in the
mass-luminosity (M-L), $M_{bol}$-mass (M-M), mass-radius (M-R) and
H-R diagrams (the diagrams are displayed in figures 6 and 7, it
appears that $EE$ $Aqr$ contains two main-sequence stars with
F2IV+K4III spectral type while the primary of $Z$ $Vul$ is still a
main-sequence object, and the secondary component is on its way to
the giant stage with B2V+B9V spectral type for the primary and
secondary respectively (according to tables of Straizys \&
Kuriliene, 1981). The results are tabulated in table 10 and compared
with the results of the pervious authors.

\begin{table*}
\caption{Other elements of the binary system $EE$ $Aqr$}
\begin{tabular}{@{}llllll@{}}
\hline
Parameters & This Work & This Work & This Work & Hilditch and \\
& Filter B & Filter U & Filter V & King (1988) \\
& (Johnson) & (Johnson) & (Johnson) & \\
\hline
$a_1\sin i$ & 0.9876 & 0.9899 & 0.9827 & 0.92 \\
$a_2\sin i$ & 3.0880 & 3.0951 & 3.0728 & 2.85 \\
$K_1 (Km/s)$ & 98.2403 & 98.4678 & 97.7579 & 91.3 \\
$K_2 (Km/s)$ & 307.1819 & 307.8929 & 305.6733 & 284 \\
$fillout$ $1$(\%) & 95.882 & 96.020 & 96.428 & 86 \\
$fillout$ $2$(\%) & 100.000  & 100.000 & 100.000 & 94 \\
\hline
\end{tabular}
\end{table*}

\begin{table*}
\caption{Other elements of the binary system $Z$ $Vul$}
\begin{tabular}{@{}llllll@{}}
\hline
Parameters & This Work & This Work & This Work & Plaskett (1920) & Popper (1957b) \\
& Filter B & Filter U & Filter V & & \\
& (Johnson) & (Johnson) & (Johnson) & & \\
\hline
$a_1\sin i$ & 4.7716 & 4.7714 & 4.7717 & & \\
$a_2\sin i$ & 11.0963 & 11.0958 & 11.0964 & & \\
$K_1 (Km/s)$ & 98.4167 & 98.4127 & 98.4182 & 96.4 & 89.8 \\
$K_2 (Km/s)$ & 228.8649 & 228.8557 & 228.8685 & 213.7 & 219.7 \\
$fillout$ $1$(\%) & 72.118 & 71.112 & 73.161 & & \\
$fillout$ $2$(\%) & 100.000 & 100.000 & 100.000 & & \\
\hline
\end{tabular}
\end{table*}

\begin{figure}[t]
\includegraphics[width=0.48\textwidth]{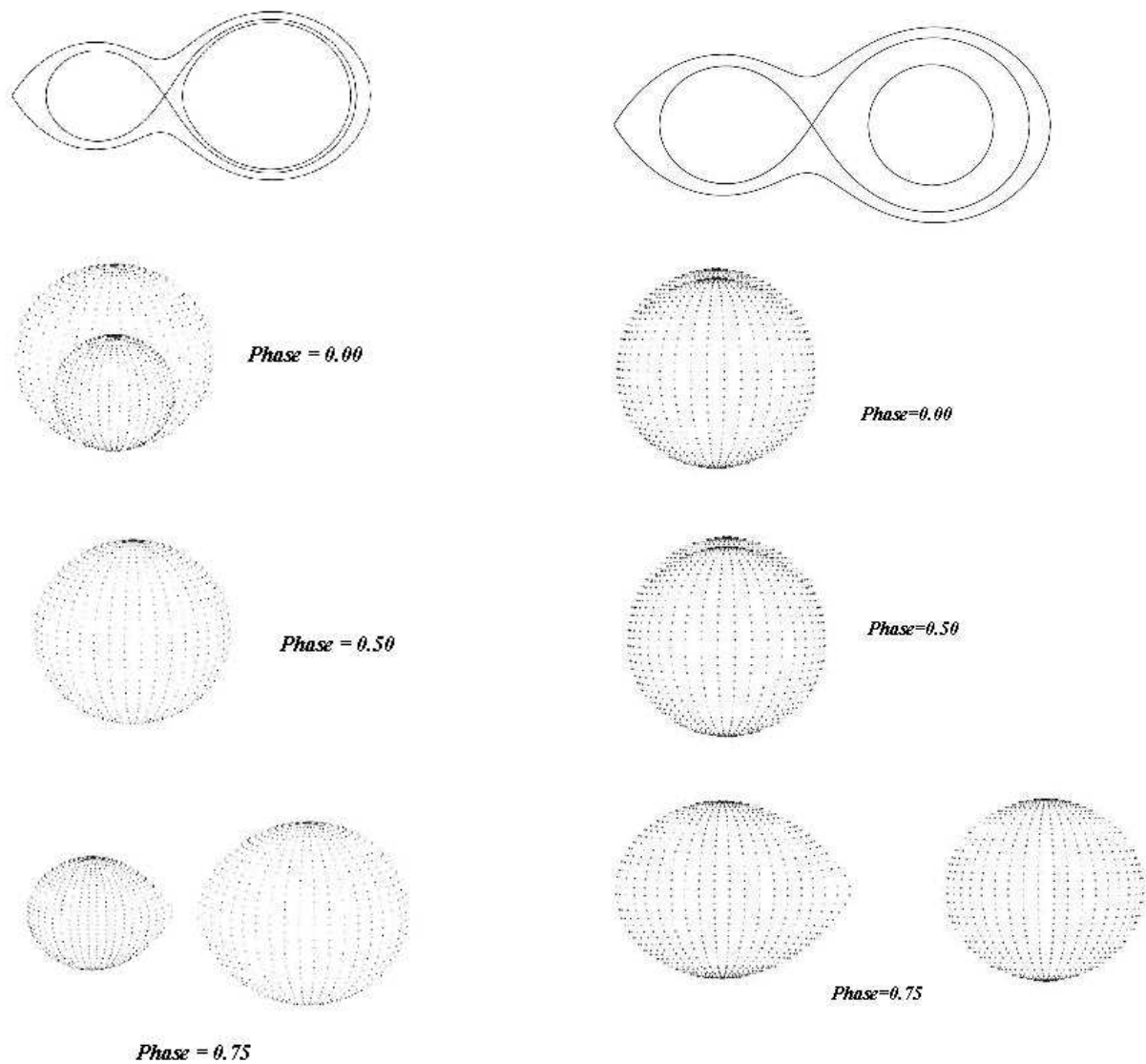}
\caption{Configuration of binary system $EE$ $Aqr$}
\end{figure}

\begin{figure}[t]
\includegraphics[width=0.48\textwidth]{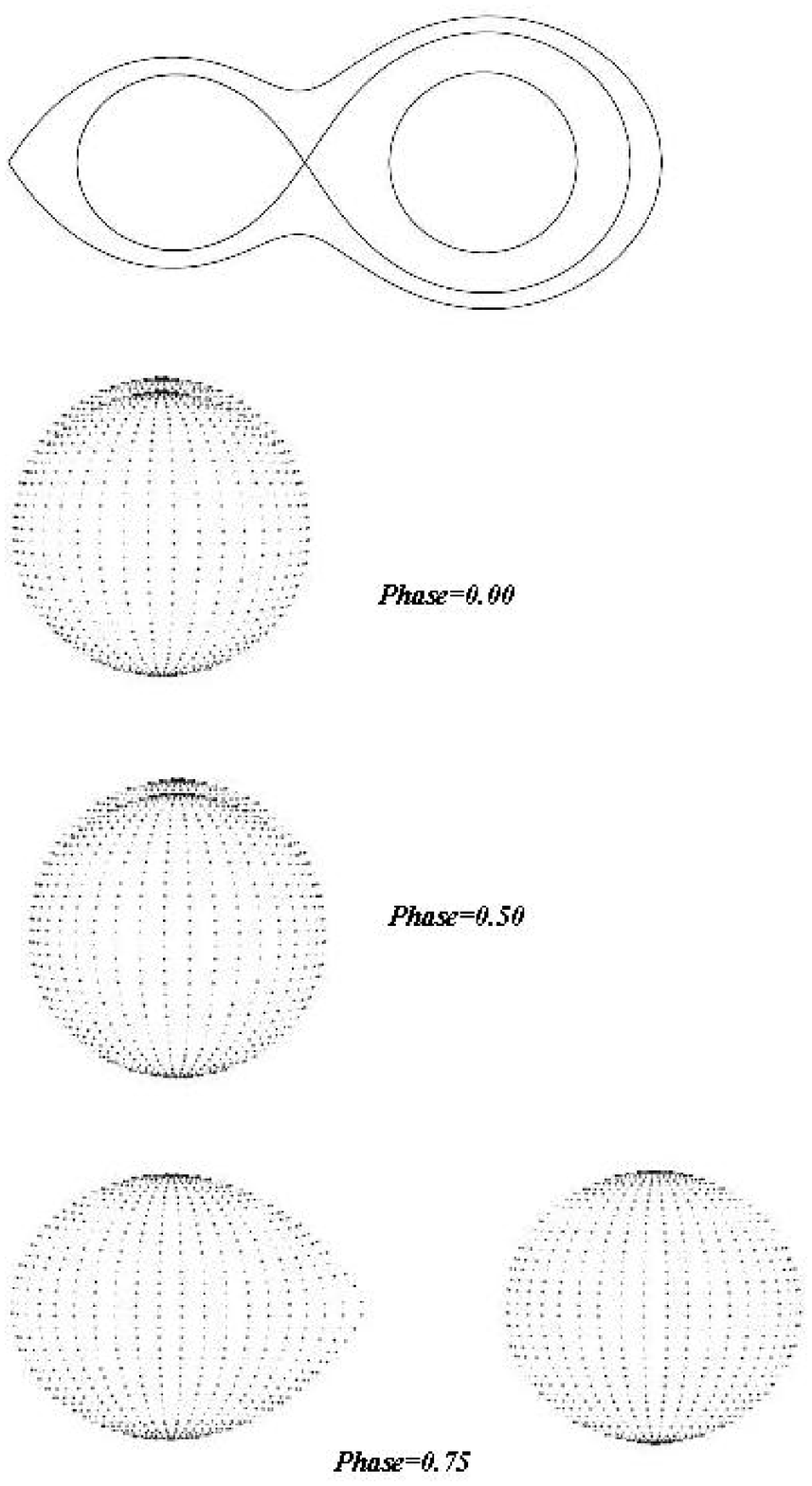}
\caption{Configuration of binary system $Z$ $Vul$}
\end{figure}

\begin{table*}
\small
\caption{Spectral type of $EE$ $Aqr$ and $Z$ $Vul$}
\begin{tabular}{@{}lllllllllll@{}}
\hline
Star & This work & Williamon & Padalia & Russo and & Plaskett & Petrie & Roman & Poper & Cester et al. \\
& & (1974) & (1979) & Sollazzo & (1920) & (1950) & (1956)& (1957b) & (1977) \\
& & & & (1982) & & &  & & \\
\hline
$EE$ $Aqr$ (primary) & F2 IV & F0 & F V & A8 V & & & & & \\
$EE$ $Aqr$ (secondary) & K4 III & & A & K3-K4 & & & & & \\
$Z$ $Vul$ (primary) & B2 V & & & & B3 & B4 & B5 V & B3-4 V & B2 V \\
$Z$ $Vul$ (secondary) & B9 V & & & & B3 & B6 & A & A2-3 III & A1 III \\
\hline
\end{tabular}
\end{table*}

\begin{figure}[t]
\includegraphics[width=0.48\textwidth]{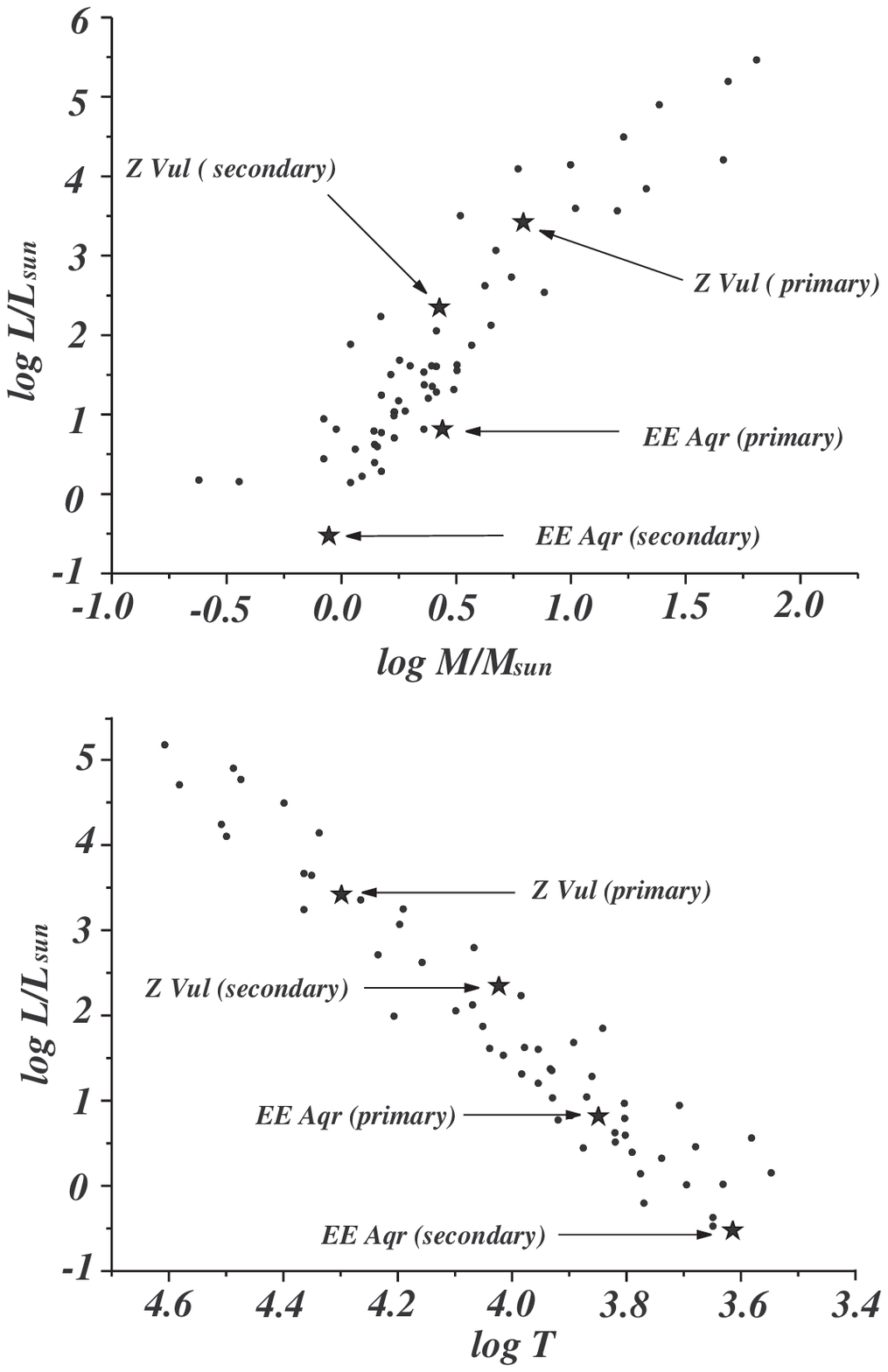}
\caption{M-L and H-R diagram. The location of components are shown
with star sign.}
\end{figure}

\begin{figure}[t]
\includegraphics[width=0.48\textwidth]{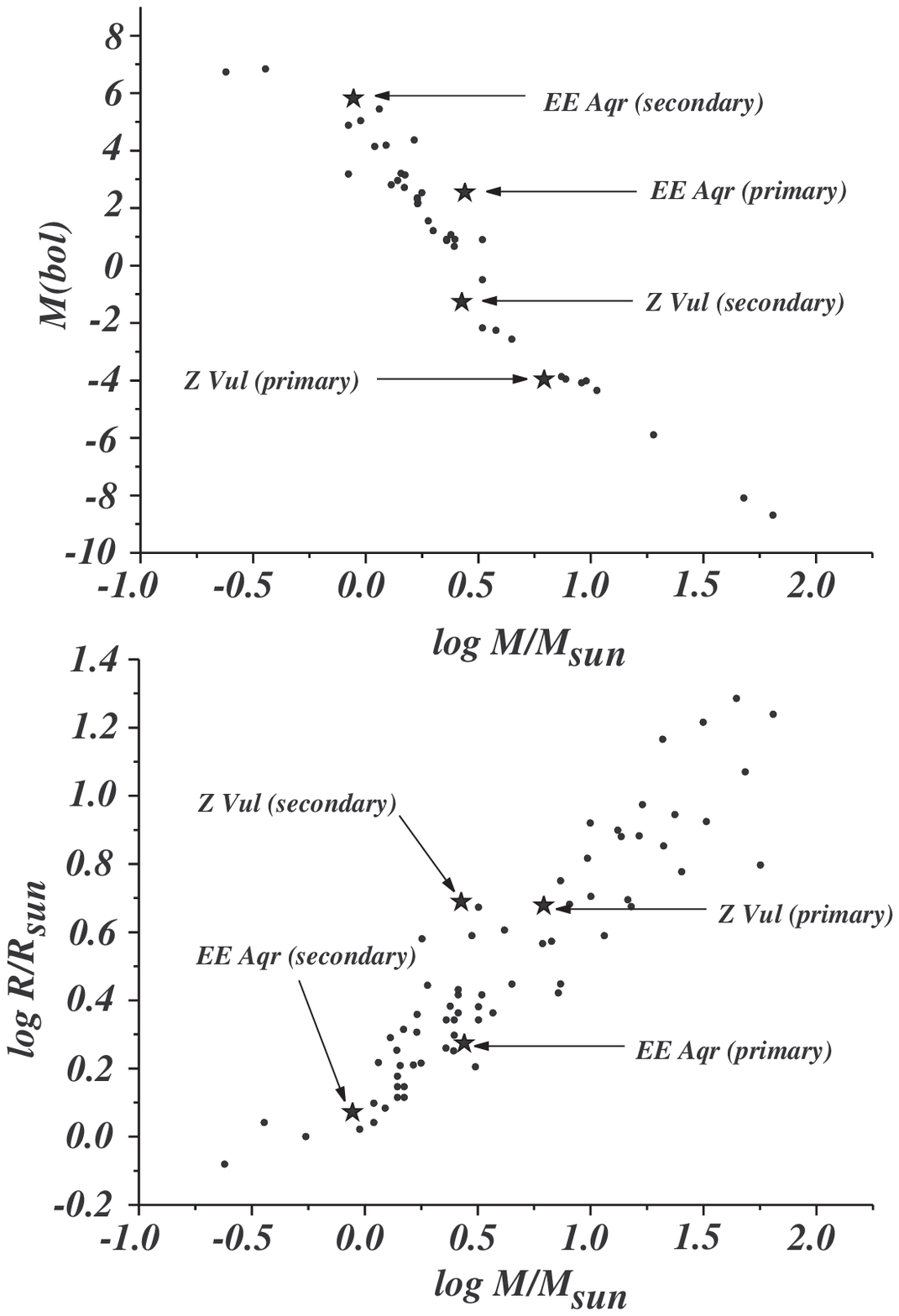}
\caption{M-M and M-R diagram. The location of components are shown
with star sign.}
\end{figure}


\acknowledgments We wish to very thank professor Bahram Khalesseh
for his useful comments and discussion at various stage of the work.
Also we would like to thank the referee for their comments and
suggestions.


\end{document}